# MAKING NUCLEI OUT OF THE SKYRME CRYSTAL

W.K. Baskerville[1]

Department of Applied Mathematics
& Theoretical Physics
University of Cambridge
Cambridge CB3 9EW, U.K.

March 1995

**ABSTRACT:** A new method for approximating Skyrme solutions is developed. It consists of cutting sections out of the Skyrme crystal and smoothly interpolating between the boundary and spatial infinity. Several field configurations are constructed, and their energies calculated. The surface energy (per unit area) of an infinite flat plane of the crystal is also calculated, and the result used to derive a formula analogous to the semi-empirical mass formula of nuclear physics. This formula can be used to give some idea of what the Skyrme model predicts about volume and surface energies of the nucleus over a broad range of baryon numbers.

## 1  Introduction.

The Skyrme model [1] has had considerable qualitative success in describing nucleons as solitons in a non-linear field theory of $\pi$-mesons. It is known that in the large-$N$ limit, QCD reduces to an effective meson field theory [2]. The Skyrme model provides a particularly simple example, since it includes only pions. It can be thought of as a low-energy approximation to a more complicated theory with more mesons.

Interest in the theory first arose when Skyrme demonstrated that by adding an extra fourth order term to the Lagrangian of the non-linear sigma model, it is possible to ensure the existence of soliton solutions to the field equations [1]. His remarkable suggestion that these solitons might be baryons was prompted by the existence of a conserved topological charge, which he speculated could possibly be identified with the baryon number ($B$) of the field. This conjecture was confirmed by Witten in 1983 [3], who showed that the Skyrme model solitons, when quantised, reproduce exactly the quantum numbers of QCD baryons. Thus $B = 1$ skyrmions can be interpreted as classical states of nucleons, deltas and higher resonances. Similarly, multisoliton solutions with $B > 1$ model classical states of larger nuclei.

---

[1] Work supported by Packer Australia Scholarship (Cambridge Commonwealth Trust), and Overseas Research Studentship.

The minimal energy solutions for the first few baryon numbers (up to $B = 6$), have been calculated in detail numerically [4]. Interestingly, these solutions all display very definite symmetries. A single skyrmion has spherically symmetric energy density, while the $B = 2$, $B = 3$ and $B = 4$ solutions display axial, tetrahedral and cubic symmetries respectively. The latter is particularly interesting since its energy per baryon is noticably lower than all the other solutions thus far calculated. This is a good sign if we hope to identify it with (a classical version of) the alpha particle. The $B = 5$ and $B = 6$ solutions have less symmetry.

These finite baryon number solutions are not the only stable configurations possible in Skyrme theory. Infinite periodic solutions also exist, of which that of lowest energy per baryon is called the Skyrme crystal. This consists of half-skyrmions arranged on a simple cubic lattice [5]. The energy per baryon of this crystal is only four percent above the topological lower bound.

The present research begins with the observation that there is a remarkable similarity between the energy distribution in the unit cell of this crystal, and the $B = 4$ solution. Might it be possible to approximate finite nuclei by cutting sections out of the crystal, and then smoothly interpolating between the crystal boundary and spatial infinity? The $B = 4$ solution is an obvious candidate for such a procedure, but the approximation should actually improve as $B$ increases, since surface effects arising from cutting the crystal should have less impact on larger sections, where the volume energy is relatively greater. It is just such large-$B$ solutions which have not been modelled by any other method.

Another attractive feature of this model arises from the fact that the crystal fields can be extremely well approximated by analytic formulae. It is also convenient to express the outside fields analytically. This greatly facilitates computation, as well as potentially providing greater insight into the solutions obtained. A computer is necessary only for performing integrations, unlike most other Skyrme calculations, which rely heavily on numerics.

Some general calculational details will be discussed in section 3, after a brief review of the Skyrme model in Section 2. First the Skyrme crystal is described in Section 3.1. The ansatz adopted for the outside region is then explained in Section 3.2. The energy calculations involve cutting a section out of the crystal, and smoothly interpolating between the boundary and spatial infinity. For a configuration with finite potential energy, there is an arbitrary choice of the 'vacuum' at spatial infinity. It is shown that an $O(4)$ rotation must be performed on the crystal fields, to ensure compatibility between the crystal symmetry and this boundary condition. Section 3.3 concludes with a brief description of the properties of the crystal after this rotation has been made .

In Section 4, a brief digression is made from the main topic to discuss a related problem, namely the calculation of the surface energy (per unit area) of the crystal. The approximations necessary to simulate an infinite surface are discussed in Section 4.1. The asymptotic behaviour of the outside ansatz, and its relation to the geometry, are then discussed in detail in Section 4.2, after which the results of the calculation are given in Section 4.3. Finally, an improvement to the ansatz (consisting of using curved rather than flat boundaries) is considered in Section 4.4.

To conclude, the results of the main calculations are given and discussed in Section 5. A discussion of the type of sections which can sensibly be cut from the crystal is given in Section 5.1, where it will be shown that simple cubes are the most energetically favourable shapes. A few calculational shortcuts for this case will then be given in Section 5.2, followed by the results for simple cubes in Section 5.3. Finally, other shapes with cubic symmetry are briefly considered in Section 5.4. An energy formula, comparable to the semi-empirical mass formula of nuclear physics, is derived for simple cubes. It is suggested that this formula may be valid for a greater range of baryon numbers than those strictly corresponding to cubic crystal sections.

## 2    The Skyrme Model.

The Skyrme field is a smooth, scalar $SU(2)$-valued field $U(x_0, \underline{x})$, which may be written

$$U(x_0, \underline{x}) = \sigma(x_0, \underline{x}) + i\underline{\pi}(x_0, \underline{x}).\underline{\tau} \qquad (1)$$

where

$$\sigma^2 + \pi_1{}^2 + \pi_2{}^2 + \pi_3{}^2 = 1. \qquad (2)$$

$\pi_1$, $\pi_2$ and $\pi_3$ are to be identified as a triplet of pion fields. In terms of the Lie algebra valued current $L_\mu = U^{-1} \partial_\mu U$ ($\mu = 0, 1, 2, 3$), the Skyrme Lagrangian density is [6]

$$\mathcal{L} = \frac{F_\pi^2}{16} Tr(L_\mu L^\mu) + \frac{1}{32e^2} Tr([L_\mu, L_\nu][L^\mu, L^\nu]). \qquad (3)$$

Here [ , ] denotes the bracket of the $SU(2)$ Lie algebra, and a Minkowski metric $(-, +, +, +)$ is assumed. The constants $F_\pi$ and $e$ are free parameters of the Skyrme model, generally determined by appropriate comparison with experiment. It is convenient to work with a rescaled Lagrangian, in which they are absent:

$$\mathcal{L} = \frac{1}{2} Tr(L_\mu L^\mu) + \frac{1}{16} Tr([L_\mu, L_\nu][L^\mu, L^\nu]). \qquad (4)$$

This is done by taking the unit of energy to be $F_\pi/4e$, and the unit of length as $2/F_\pi e$. We follow standard practice by adopting the approach of [7, 8], where $F_\pi$ and $e$ are chosen to fit the masses of the nucleon and the delta resonance, with and without the physical pion mass respectively. Since the pion mass is here assumed to be zero, we adopt the latter convention. Our units are thus related to conventional units as follows:

$$\frac{F_\pi}{4e} = 5.92 \text{ MeV}, \qquad \frac{2}{F_\pi e} = 0.561 \text{ fm}. \qquad (5)$$

In these units, the potential energy of a Skyrme field at a given time is

$$E = \int d^3x \left( -\frac{1}{2} Tr(L_i L_i) - \frac{1}{16} Tr([L_i, L_j][L_i, L_j]) \right) \quad (6)$$

($i, j = 1, 2, 3$). If the potential energy is finite, $U$ must tend to a constant value at infinity, independent of direction. The conventional choice is $U \to 1$. Once this condition is satisfied, the baryon number of the field can be written

$$B = -\frac{1}{24\pi^2} \int d^3x\, \epsilon_{ijk} Tr(L_i L_j L_k). \quad (7)$$

$B$ is a winding number: the degree of the map $U(\underline{x}) : R^3 \to S^3$. Essentially, it is the number of times each possible field value occurs. The energy satisfies the topological constraint

$$E \geq 12\pi^2 |B| \quad (8)$$

and it is usual to express the energy of a given configuration in terms of this bound. The notation $E^*$ will be used when quoting energies in this way (ie. $E^* = E/(12\pi^2 |B|)$).

It is sometimes more convenient to use equivalent forms of these expressions for $E$ and $B$ in actual calculations. We adopt standard conventions and define the components $\underline{a}_\mu$ of $L_\mu$ by

$$L_\mu(\underline{x}) = i\underline{a}_\mu(\underline{x}).\underline{\tau}. \quad (9)$$

Explicitly

$$\underline{a}_\mu = \underline{\pi} \wedge \partial_\mu \underline{\pi} - \underline{\pi} \partial_\mu \sigma + \sigma \partial_\mu \underline{\pi}. \quad (10)$$

The baryon and energy densities may then be written in terms of $\underline{a}_i$ as

$$\mathcal{B}(\underline{x}) = -\left(\frac{1}{2\pi^2}\right) \underline{a}_1(\underline{x}).\underline{a}_2(\underline{x}) \wedge \underline{a}_3(\underline{x}). \quad (11)$$

$$\mathcal{E}(\underline{x}) = \underline{a}_i.\underline{a}_i + \frac{1}{2}\left((\underline{a}_i.\underline{a}_i)(\underline{a}_j.\underline{a}_j) - (\underline{a}_i.\underline{a}_j)(\underline{a}_i.\underline{a}_j)\right). \quad (12)$$

The energy density may also be written directly in terms of the field derivatives, by noting that $\underline{a}_i.\underline{a}_j = \partial_i \underline{\pi}.\partial_j \underline{\pi} + \partial_i \sigma \partial_j \sigma$.

## 3  General Calculational Details.

In this Section some details common to both the infinite plane and the finite crystal section calculations will be described. In both cases, all calculations of physical quantities split naturally into two parts, dealing with the regions inside and outside the crystal respectively. We will therefore begin with a discussion of the Skyrme crystal, and then move on to the ansatz employed in the outside region when the crystal is cut.

## 3.1 The Skyrme Crystal.

The Skyrme crystal was discovered by Kugler and Shtrikman in 1988 [5], and also independently by Castillejo et al. [9]. It consists of half-skyrmions arranged on a simple cubic lattice.

The concept of a "half-skyrmion" may at first seem a little strange, but is actually quite well-defined. A single skyrmion is spherically symmetric, as was previously mentioned. It is conventional to take $\sigma = 1$ ($\pi_i = 0$) as the field value at spatial infinity. With this choice, $\sigma = -1$ at the centre of the skyrmion. Between these extremes, there is a spherical surface where $\sigma = 0$: it turns out that exactly half of the baryon density lies on each side of this surface. In the crystal, this $\sigma = 0$ surface is deformed into a cube, and these cubes are then packed in an obvious way, with the sign of $\sigma$ alternating at the centres of neighbouring half-skyrmions.

Kugler and Shtrikman [5] found the crystal fields by writing down the most general Fourier expansions consistent with the symmetry, and then determining the coefficients numerically by minimising the energy. The Skyrme crystal is thus in a sense *defined* by its symmetry.

The full cubic point group consists of forty eight elements, which can be divided into ten equivalence classes. Each class corresponds to a particular physical symmetry of a cube. Associated to each element $g$ of the cubic group, there is a linear transformation of the fields $\mathcal{D}(g)$, where $\mathcal{D}(g)$ is a $4 \times 4$ matrix. The matrices $\mathcal{D}(g)$ define a 4-dimensional representation of the group. To specify the field symmetries it is only necessary to state how the fields change under the operation of three generators of the cubic group: a reflection, a 120° rotation and a 90° rotation. All other transformations consist of some combination of these three. Explicitly, the crystal symmetry is

$$\begin{aligned}
\text{when} \quad & (x,y,z) \longrightarrow (-x,y,z) &, \quad & (\sigma, \pi_1, \pi_2, \pi_3) \longrightarrow (-\sigma, \pi_1, \pi_2, \pi_3) \\
\text{when} \quad & (x,y,z) \longrightarrow (-z,x,-y) &, \quad & (\sigma, \pi_1, \pi_2, \pi_3) \longrightarrow (\sigma, \pi_3, \pi_1, \pi_2) \\
\text{when} \quad & (x,y,z) \longrightarrow (y,-x,z) &, \quad & (\sigma, \pi_1, \pi_2, \pi_3) \longrightarrow (-\sigma, \pi_2, \pi_1, \pi_3).
\end{aligned} \quad (13)$$

The cubic group has ten irreducible representations (irreps), four of which are 1-dimensional, two 2-dimensional and four 3-dimensional. The representation formed by the crystal symmetry is 4-dimensional, and must therefore be reducible. Its decomposition into irreps can most easily be performed by a consideration of its characters. The character of a given element is equal to the trace of the matrix representing it. It is a class characteristic. A list of the characters for all classes is sufficient to uniquely identify any representation. For a full discussion of the cubic group, its classes and irreps, and its character table, see for example [10]. In fact the 4-dimensional representation associated with the crystal can be decomposed into two 1-dimensional irreps (one of which is the trivial representation) and a 2-dimensional irrep. The inclusion of the trivial 1-dimensional irrep is the crucial property which (as will be seen later) allows sections to be cut from the crystal.

The Skyrme crystal has also been studied by Castillejo et al. [9], who discovered that right at the energy minimum, the crystal fields are extremely well approximated by

$$\begin{aligned} \sigma &= \sin\alpha \sin\beta \sin\gamma \\ \pi_1 &= \cos\alpha \sqrt{1 - \frac{1}{2}\cos^2\beta - \frac{1}{2}\cos^2\gamma + \frac{1}{3}\cos^2\beta \cos^2\gamma} \end{aligned} \quad (14)$$

and cyclically for $\pi_2$ and $\pi_3$. $\alpha = \frac{\pi x}{L}$, $\beta = \frac{\pi y}{L}$ and $\gamma = \frac{\pi z}{L}$, where $L$ is the lattice parameter: the distance between the centres of neighbouring half-skyrmions. These formulae were derived by taking a three dimensional analogue of an exact two dimensional solution for the non linear $\sigma$ model. The simple analytic form of these expressions makes their adoption very convenient, and a quick check of the energy per baryon calculated using them reveals the approximation to be very good indeed: the energy per baryon found by Kugler and Shtrikman is reproduced exactly to as many decimal places as quoted, $1.038 \times 12\pi^2$.

It will be noticed that an unspecified parameter has been introduced into the expressions for the fields: the lattice parameter $L$. In fact, there is a freedom of length scale in most Skyrme theory calculations; in an infinite periodic solution this naturally takes the form of a lattice parameter. It must be determined by a minimisation of the energy of a cube of side $L$. The energy has two terms (which will henceforth be denoted $E_2$ and $E_4$), with quadratic and quartic dependence on the derivatives respectively. Obviously, $E_2 \sim L$ while $E_4 \sim 1/L$. Now, the energy will be minimised with respect to $L$ at the scale where these two terms are equal. The true energy can thus be written

$$E = 2\sqrt{E_2 E_4} \quad (15)$$

for $E_2$ and $E_4$ calculated using any, *arbitrarily chosen* $L$. We thus choose $L = \pi$, since for this value $\alpha$, $\beta$ and $\gamma$ in (14) above reduce to $x$, $y$ and $z$. The energy is then normalised to the right scale using the 'trick' above. The true lattice parameter can also be recovered:

$$L = \pi \sqrt{\frac{E_4}{E_2}}. \quad (16)$$

Note that, for a finite crystal section, the total energies $E_2$ and $E_4$ must be calculated separately for the inside and outside regions, and the *sums* used in the formulae above.

### 3.2 The Outside Ansatz.

We turn now to the construction of the fields outside the crystal. What is needed is an interpolation between the fields on the surface of the crystal section, and the vacuum ($\sigma = 1$, $\pi_i = 0$) at spatial infinity. It is necessary to ensure that the condition (2) is always fulfilled, so that the baryon number will come out correctly. We thus start with a standard parametrisation of $SU(2)$, and then insert an appropriate radial (or vertical) dependence into some of the parameters so that the fields tend to the right limit at infinity:

$$\sigma = \cos f$$
$$\pi_1 = \sin f \cos b$$
$$\pi_2 = \sin f \sin b \cos c$$
$$\pi_3 = \sin f \sin b \sin c \qquad (17)$$

where $f = ga$. The functions $a$, $b$ and $c$ are determined by the boundary conditions: they must be matched to the crystal fields on the surface. The function $g$ extrapolates the fields between the boundary and infinity. It depends either on $r$ or $z$ depending on the geometry, and can be chosen fairly arbitrarily, as long as it is monotonic, and

$$g = 1 \text{ at the boundary}, \qquad g \to 0 \text{ when } r \to \infty \text{ (or } z \to \infty\text{).} \qquad (18)$$

This ansatz has been chosen so that the interpolation between the boundary and the vacuum is as direct as possible. The field values on the crystal boundary map out a two-dimensional surface in the parameter space of $SU(2)$, which itself is a three-sphere. Figure 1 shows an analogue of this on $S^2$ ($S^3$ being too difficult to visualise). The boundary shown in this figure only loops once around the equator, for simplicity's sake, but a real boundary could wind around it several times. From any point on this boundary, the shortest path to the vacuum (taken to be the "North Pole") is a great circle, as shown. Furthermore, such a path is unique, *except* when interpolating from the point diametrically opposite the vacuum, in this case the "South Pole". It is clear that no discontinuities are introduced into the fields if this 'great circle' interpolation is used, as long as the "South Pole" is excluded from the boundary. The actual ansatz is a direct generalisation of this model onto the three-sphere. The 'bad point' which must not occur on the boundary is $\sigma = -1$ ($\pi_i = 0$). The $SU(2)$ parametrisation (17) thus not only ensures that the field values remain on the surface of $S^3$, but picks out the 'great circle' trajectories from the boundary to the vacuum.

The question now arises as to whether the imposition of the boundary condition at spatial infinity, necessary for finite crystal sections, is compatible with the crystal symmetry. From the decomposition into irreps of the 4-dimensional representation corresponding to the crystal symmetry, it is clear that one component of the $SU(2)$ field transforms trivially under the cubic group. This component must be identified with $\sigma$ if the crystal symmetry is to be compatible with the boundary condition at spatial infinity, since $\sigma$ at infinity is unaffected by any rotation or reflection. This condition is not met in the particular realisation of the crystal symmetry given in Equation (13). An $O(4)$ rotation of the crystal fields given in Equation (14) is therefore necessary, to ensure that they *do* meet this condition.

The $O(4)$ rotation required is of course not unique, since the system remains invariant under $SO(3)$ rotations involving the pion fields only. The simplest possibility is to swap $\sigma$ with whatever combination of pion fields corresponds to the trivial irrep in (13). (To maintain the sign of the baryon density, a 90° rotation should be used, rather than a reflection.) By inspection, the combination $(\pi_1 + \pi_2 + \pi_3)$ is invariant under the crystal point group. The required transformation is thus a 90° rotation in the $(\sigma, (\pi_1 + \pi_2 + \pi_3))$ plane. Explicitly, the transformed fields are

$$\widetilde{\sigma} = \pm \frac{1}{\sqrt{3}}(\pi_1 + \pi_2 + \pi_3)$$
$$\widetilde{\pi}_1 = \mp \frac{1}{\sqrt{3}}\sigma + \frac{1}{3}(2\pi_1 - \pi_2 - \pi_3) \quad \text{and cyclically.} \tag{19}$$

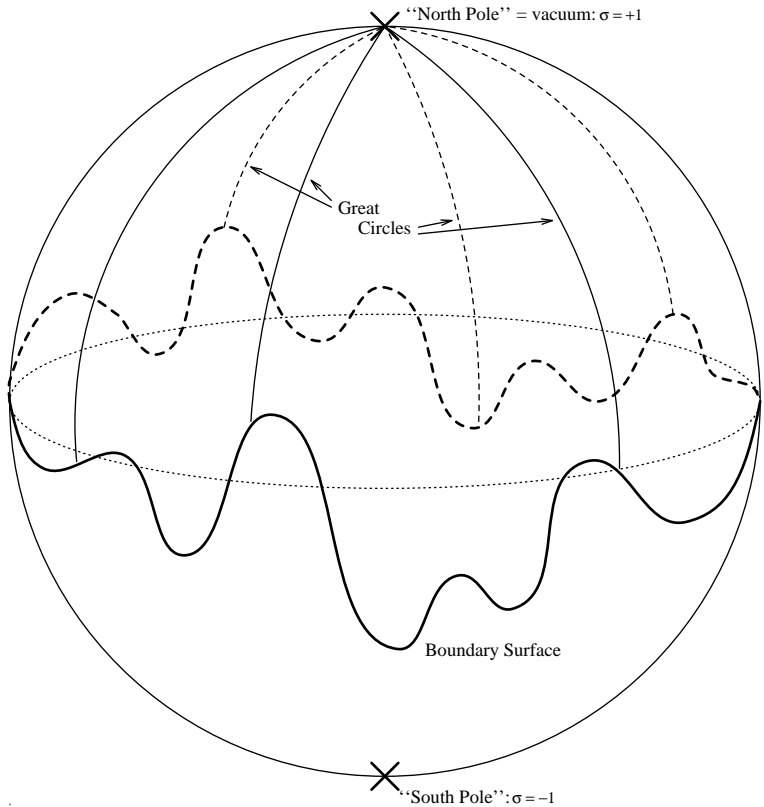

Figure 1: $S^2$ Analogue of Outside Ansatz: Trajectories of Fields from the Crystal Boundary to the Vacuum.

The ambiguity in sign is due to the possibility of rotating 90° in *either* direction. This sign is quite important: as will be explained later, it affects the baryon number of all solutions cut from the crystal. From here on we drop the ~, and assume the crystal fields have been rotated.

Note that the unrotated fields (14) may still be used directly in the calculation of all physical quantities, like the energy density, inside the crystal. However, the functions $a$, $b$ and $c$ which appear in the outside ansatz *must* be determined by matching with the rotated crystal fields (19) on the surface.

### 3.3 The 'Rotated' Crystal.

The Skyrme crystal field acquires certain definite properties when $\sigma$ is the component transforming trivially under the cubic group. We conclude this general discussion with a brief description of the most important of these for our model. The definition of the crystal unit cell will also be given.

Once $\sigma$ has been assigned to the trivial irrep, the distribution of points where $\sigma = \pm 1$ in the crystal becomes fixed. This has the interesting consequence that the baryon number of any section cut from the crystal can be calculated by a very simple formula

$$B = 4n \qquad (20)$$

where $n$ is the number of times the point $\sigma = -1$ occurs in the crystal section. (It is assumed here that $\sigma = -1$ at no points outside the crystal section.) This result is related to the fact that $\sigma = \pm 1$ at the origin, about which the fields display a four-fold symmetry, since they are invariant under a 180° rotation about any axis.

To be more explicit, the distribution of $\sigma = \pm 1$ points in the crystal is as follows: if $\sigma = -1$ at the origin, then other $\sigma = -1$ points occur at $(\pm mL, \pm nL, \pm pL)$ where $m$, $n$ and $p$ are even integers; $\sigma = +1$ points occur at $(\pm rL, \pm sL, \pm tL)$ where $r$, $s$ and $t$ are odd integers. If $\sigma = +1$ at the origin then the pattern reverses.

Let us now define the unit cell of the crystal to be a cube of side length $2L$, centred about a point where $\sigma = -1$. It is clear that the central $\sigma = -1$ point is the only one in each unit cell, while $\sigma = +1$ points occur at all the corners, (each one being shared with seven other unit cells). The baryon number of a unit cell is exactly four. Since $\sigma = \pm 1$ at the origin, the latter must be situated either at the centre of a unit cell, or at a point where the corners of eight unit cells meet. The significance of the variable sign in transformation (19) now becomes clear: it allows us to choose between these two possibilities.

## 4 The Infinite Plane: Calculating the Surface Energy of the Crystal.

Before moving on to the main discussion of the sections cut from the crystal, we will first consider a related calculation, that of the surface energy of an infinite plane of the crystal. There are two reasons for considering the calculations in this order. The first is that the discussion of asymptotics necessary to understand the form of the fall-off in the outside field flows more naturally if the infinite plane is used as the starting point. The second is that the surface energy, once calculated, can be used to derive a 'mass formula', the predictions of which can then be compared to the results for finite crystal sections. The half crystal is also an interesting problem in itself.

## 4.1 Calculational Details.

To simulate an infinite half crystal, the extrapolation outside the crystal is vertical. Also, the lattice parameter $L$ is held fixed at the value it assumes in the infinite crystal.

We consider a semi-infinite column of crystal, with a square cross-section of area $L \times L$. The crystal boundary is taken to be the plane $z = k = hL$. The functions $a$, $b$ and $c$ which appear in the outside ansatz (17) thus depend on $x$ and $y$

$$
\begin{aligned}
a &= \arccos\left(\sigma(x,y,k)\right) \\
b &= \arccos\left(\frac{\pi_1(x,y,k)}{\sin a}\right) \\
c' &= \arccos\left(\frac{\pi_2(x,y,k)}{\sin a \sin b}\right) \\
c &= \begin{cases} c' & \pi_3(x,y,k) \geq 0 \\ 2\pi - c' & \pi_3(x,y,k) < 0 \end{cases},
\end{aligned}
\tag{21}
$$

where $\arccos(\mu)$ lies in the range $[0, \pi]$. The constant $k$ (or $h$) must be determined by energy minimisation.

An exponential falloff is used for the field outside the crystal. The geometry of the column makes this appropriate, as will be explained more fully in Section 4.2 below. Explicitly,

$$g(z) = e^{-\frac{\beta}{L}(z-k)}. \tag{22}$$

The parameter $\beta$ is also to be determined by energy minimisation.

## 4.2 The Geometry Of The Column And The Asymptotic Fall-off.

Effectively, the geometry of the crystal gives rise to "mass terms" in $\mathcal{L}$, with the well-known consequence that the fall-off becomes exponential.

The full expression for the potential energy of the column (outside the crystal) is rather complicated. However, only the quadratic term of the energy density will contribute to the asymptotic limit $f \longrightarrow 0$, $z \longrightarrow \infty$.

$$
\begin{aligned}
\mathcal{E}_2 &= a^2\left(\frac{df}{dz}\right)^2 + \left(\left(\frac{da}{dx}\right)^2 + \left(\frac{da}{dy}\right)^2\right) f^2 \\
&\quad + \left\{\left(\frac{db}{dx}\right)^2 + \left(\frac{db}{dy}\right)^2 + \left(\left(\frac{dc}{dx}\right)^2 + \left(\frac{dc}{dy}\right)^2\right)\sin^2 b\right\}\sin^2 af
\end{aligned}
\tag{23}
$$

It is not possible to integrate out the $x$ and $y$ dependence before taking the asymptotic limit in this case. However, once the limit has been taken, separation becomes possible, and the energy density can then be expressed in terms of $z$ as follows

$$\mathcal{E} \longrightarrow A\left(\frac{df}{dz}\right)^2 + Bf^2 \tag{24}$$

where

$$A = \int_0^L \int_0^L a(x,y)^2 \, dx \, dy$$

$$B = \int_0^L \int_0^L \left( \left(\frac{da}{dx}\right)^2 + \left(\frac{da}{dy}\right)^2 \right.$$

$$\left. + a^2 \left\{ \left(\frac{db}{dx}\right)^2 + \left(\frac{db}{dy}\right)^2 + \left( \left(\frac{dc}{dx}\right)^2 + \left(\frac{dc}{dy}\right)^2 \right) \sin^2 b \right\} \right) dx \, dy. \quad (25)$$

Since the functions $a, b$ and $c$ depend on the boundary conditions, the integrals $A$ and $B$ depend on $k$, the $z$-coordinate of the cut. However, they have no explicit dependence on $z$. The lack of any z-dependence in the derivative term means that the quadratic term in $f$ can now act as a mass term.

Compare Equation (24) to the energy density (integrated over angles) of the well-known hedgehog solution ($U(\underline{x}) = \exp i f(r) \hat{\underline{x}}.\underline{\tau}$), in the asymptotic limit

$$\mathcal{E} \longrightarrow r^2 \left(\frac{df}{dr}\right)^2 + 2f^2. \quad (26)$$

This gives the asymptotic field equation

$$\frac{d^2 f}{dr^2} + \frac{2}{r} \frac{df}{dr} - \frac{2}{r^2} f = 0, \quad (27)$$

which leads to the well-known dipole falloff of a single skyrmion, $f \sim r^{-2}$. Now consider the asymptotic field equation derived from the energy expression (24)

$$\frac{d^2 f}{dz^2} = \frac{B}{A} f, \quad (28)$$

which gives the solution

$$f \sim e^{-\sqrt{\frac{B}{A}} z}. \quad (29)$$

It can thus quite clearly be seen how the geometry of the column leads to an exponential tail in the asymptotic limit.

### 4.3 Results.

It is to be expected that there will be a local minimum in surface energy whenever the crystal is cut just below a plane where $z = pL$, where $p$ is an odd (even) integer if $\sigma = -1(+1)$ at the origin. Because of the translational invariance of the crystal, and of the outside ansatz (22), these minima are the same for any integer value of $p$ (assuming appropriate choices for $\sigma$ at the origin).

The precise definition of 'surface energy' ($S$) used is as follows. If $p - 1 < h < p$ then

$$S = E_T(h) - pV \quad (30)$$

| $h(x, y)$ | $E_s$ | $\beta$ | $\alpha$ | $c$ |
|---|---|---|---|---|
| $h_1$ | 5.671 | 2.08 | 0.02 | 0.84 |
| $h_2$ | 5.668 | 2.08 | 0.01 | 0.83 |
| $h_3$ | 5.671 | 2.09 | 0.01 | 0.83 |

Table 1: Calculated Surface Energies for Curved Boundaries.

where $V$ is the volume energy of a single half-skyrmion cube in the crystal, and $E_T(h)$ is the total energy (from $z = 0$ upwards) of the column cut at $z = hL$, including the energy above the cut.

The result of this calculation, using the exponential fall-off given above, is that

$$S = 5.677 \qquad (31)$$

in our units, for a surface of area $L \times L$. For comparison, the volume energy of one half-skyrmion inside the crystal is 67.476 using the same energy units. The parameters at the energy minimum are

$$\beta = 2.09, \qquad h = 0.83. \qquad (32)$$

The calculated energy is the same at $h = 1.83$, $h = 2.83$ etc., as it should be, providing a check on this result. The column can thus be interpreted as a true 'half crystal' field configuration.

## 4.4 Curved Boundaries.

One final comment: in an attempt to improve the outside ansatz, curved boundaries were considered in the column approximation. This was achieved by considering the parameter $h$ to be a periodic function of $x$ and $y$. The following functions were considered

$$h_1(x,y) = -\alpha \sin\left(\frac{\pi x}{L}\right) \sin\left(\frac{\pi y}{L}\right) + c \qquad (33)$$

$$h_2(x,y) = -\alpha \cos\left(\frac{2\pi x}{L}\right) \cos\left(\frac{2\pi y}{L}\right) + c \qquad (34)$$

$$h_3(x,y) = -\frac{\alpha}{2}\left(\cos\left(\frac{\pi x}{L}\right) + \cos\left(\frac{\pi y}{L}\right)\right) + c. \qquad (35)$$

The results for each of these choices are given in Table 1. The best curved surface lowers the surface energy by only 0.15%. The parameters which minimise the energy are also very little changed by using curved rather than flat boundaries.

We have therefore chosen not to attempt to use curved boundaries in the cubic section calculations, since the difference produced is insignificant when compared to the fairly drastic approximation involved in the whole procedure. It is reassuring, however, to know that taking the boundary to be composed of flat planes does not much affect the results (with the possible exception of edge and corner effects).

# 5 Main Results: Crystal Sections.

We turn now to the main calculations, involving finite sections cut out of the crystal. We begin by discussing which shapes can sensibly be cut from the crystal.

## 5.1 The Choice Of Crystal Section.

It will be remembered from Section 3.3 that the baryon number of a given crystal section depends only on the number of $\sigma = -1$ points contained in it. When combined with the actual distribution of such points, this leads to strong restrictions on the type of sections which can be cut from the crystal.

To preserve cubic symmetry, any section cut from the crystal must obviously be cubically symmetric about the origin. Also, to minimise the energy, most of the baryon density should be inside the crystal section rather than in the outside field. This means that the volume of a section must be carefully considered in relation to the number of $\sigma = -1$ points it contains. Furthermore, it turns out that it is also energetically unfavourable to let the boundary of a crystal section come too close to a $\sigma = -1$ point. (Remember that $\sigma = -1$ is the 'bad point' in our outside ansatz, which must not occur *on* the boundary.) Effectively, this means that if a particular $\sigma = -1$ point is included in the section, then most of the unit cell surrounding it should also be included. So the only sections that can (sensibly!) be cut from the crystal are composed of whole unit cells, arranged in a cubically symmetric way about the origin. To be more precise, the crystal must be cut just *inside* the boundaries of the outer unit cell(s) of such a section; since by definition they already have exactly the right baryon number relative to the number of $\sigma = -1$ points contained, and the outside region must also contain some baryon density.

Of all the possible cuboidal sections, the most energetically favourable shape is a simple cube, since this maximises the volume energy relative to surface energy. The smallest cube which can be cut from the crystal is just a single unit cell (centred on the origin), which will give a $B = 4$ configuration. The next ($B = 32$) consists of eight unit cells whose corners meet at the origin. Two more cubic sections give baryon numbers consistent with real nuclei: $B = 108$, composed of nine unit cells; and $B = 256$, composed of sixteen unit cells. If the section has an even number of unit cells, then $\sigma = +1$ at the origin; if it has an odd number, then $\sigma = -1$.

## 5.2 Calculational Shortcuts for Simple Cubes.

Several calculational shortcuts are possible for simple cubes. Firstly, because of the cubic symmetry, it is only necessary to consider one octant of the cube. We choose the octant where all three Cartesian axes are positive, for convenience. Furthermore, all physical properties, such as baryon number and energy densities, will be unaffected by any operation of the cubic group.

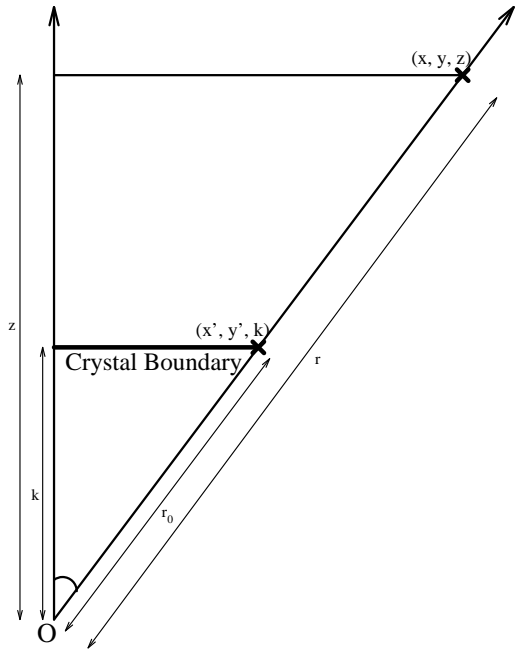

Figure 2: Relation between Cartesian coordinates at the boundary, and radially out from it.

For example, a 120° rotation will map one face of a cube into another. It is therefore sufficient to consider extrapolations along radial lines emanating only from one face of the cube: say that for which $z$ is a constant.

It is now necessary to choose a coordinate system to describe the outside ansatz. The obvious coordinates with which to describe the boundary surface are the $x$ and $y$ coordinates *at the boundary*. However, the extrapolation will be radial, which would suggest polar coordinates as more appropriate. This apparent conflict can be reconciled by the following trick. Consider a crystal boundary lying in the plane $z = k$, as shown in Figure 2. It is clear that $(r_0/r) = (k/z)$, so that $z$ can replace $r$ as the 'radial' coordinate ($r_0$ is the radial coordinate at the boundary). We can make the change of coordinates

$$x' = \frac{k}{z}x, \quad y' = \frac{k}{z}y, \quad z' = z \tag{36}$$

so that if $(x, y, z)$ corresponds to the point $(r, \theta, \phi)$, then $(x', y', k)$ corresponds to $(r_0, \theta, \phi)$. The volume element becomes

$$dx\,dy\,dz = \left(\frac{z'}{k}\right)^2 dx'\,dy'\,dz'. \tag{37}$$

This allows energy and baryon densities to be calculated according to the formulae given in Section 2; the required derivatives with respect to $x$, $y$ and $z$ are calculated by the chain rule in terms of derivatives with respect to the new coordinates $x'$, $y'$ and $z'$.

The functions $a$, $b$ and $c$, which are determined by matching with the (rotated) crystal fields at the boundary, are obviously now functions of $x'$ and $y'$. Their form is exactly as given in Equation (21) for the infinite plane, replacing $x$ and $y$ by $x'$ and $y'$ throughout.

An asymptotic analysis, such as was performed for the column, indicates that a power law falloff in the outside field should be appropriate now that the extrapolation is radial rather than vertical. The expression for the energy is rather messy (like Equation (23)), so that a fractional power law, with the power depending on the size of the section, would be expected. We therefore define

$$g(r) = \left(\frac{r_0}{r}\right)^n \quad \text{or} \quad g(z) = \left(\frac{k}{z}\right)^n. \tag{38}$$

The power $n$ is kept as a parameter, to be determined by energy minimisation.

### 5.3 Results for Simple Cubes.

The surface energy calculated in Section 4 can be used to derive a simple formula for the energy of a general cubic section. It will consist of volume and surface energy terms, analogous to the first two terms of the semi-empirical mass formula of nuclear physics.

To meet the requirement that shapes cut out of the crystal must consist of whole unit cells, a simple cube must contain $8p^3$ half-skyrmions, where $p$ is a positive integer. The baryon number of such a cube is then $4p^3$. The surface and volume energies are respectively

$$E_S = 24p^2 S \tag{39}$$
$$E_V = 8p^3 V \ (= 2BV) \tag{40}$$

where $V = 61.476$ is the volume energy of a single half skyrmion (volume $L^3$), and $S = 5.677$ is the surface energy of one face of a half-skyrmion (surface area $L^2$). The total energy per baryon then, in the standarised units $E^* = E/12\pi^2$, is

$$E^* = 1.038 + 0.288\left(\frac{1}{p}\right). \tag{41}$$

How do the predictions of this formula compare to actual results? Since it neglects edge and corner effects, it might be expected to be rather more accurate for larger cubes. This is in fact the case.

We have calculated the energies of cubic sections of the crystal, using the radial extrapolation of the fields from the crystal boundary to spatial infinity described in Section 5.2. The power $n$, lattice spacing $L$ and boundary height parameter $h$ have all been adjusted to minimise the energy. Table 2 shows the predictions and actual results for the energies of the four smallest cubes. Calculations were performed using both power law and exponential forms for the radial fall-off. A power law is indicated by asymptotic analysis for cubic sections, but since the surface energy of the mass formula was calculated using a vertical falloff (and hence exponential form), an exponential form was also used in these calculations for comparison.

| $p$ | $B$ | $E^*$(formula 41) | $E^*$(calculated) | |
|---|---|---|---|---|
| | | | Power Law | Exponential |
| 1 | 4 | 1.326 | 1.222 | 1.189 |
| 2 | 32 | 1.182 | 1.193 | 1.191 |
| 3 | 108 | 1.134 | 1.150 | 1.149 |
| 4 | 256 | 1.110 | 1.125 | 1.124 |

Table 2: Calculated and Predicted Energies for Simple Cubes.

In fact, except for $B = 4$, it makes very little difference which ansatz is used for the radial fall-off. The difference becomes more and more slight as the size of the cube increases. As for the performance of the energy formula (41), it can be seen to give much too large a value for $B = 4$. There is then a kind of "cross-over" at $B = 32$, after which the formula slightly underestimates the remaining energies.

The details of the results, including the values of the free parameters at the energy minimum, are given in Tables 3 and 4. The baryon numbers have also been calculated, as a check on the numerical accuracy of the calculations. The lattice parameter $L$ (as given by Equation (16) in Section 2) is also included.

An examination of the parameters once again shows a close similarity between the two extrapolations for the $B = 32$, $B = 108$ and $B = 256$ calculations. In particular, the lattice and cut-off parameters ($L$ and $h$) are almost identical, indicating that the field configurations are of the same size in both schemes. The lattice parameter increases with baryon number, implying that the larger cubes are less dense than the smaller. For example, inside the crystal boundary, the $B = 256$ configuration contains 0.261 baryons/fm$^3$, as compared to 0.275 baryons/fm$^3$ for $B = 108$ and 0.303 baryons/fm$^3$ for $B = 32$.

The $B = 4$ results are rather anomalous: $h$ and $L$ are extremely low, indicating that the crystal section is very small; and the exponential parameter $\beta$ is very different from that obtained for the other three configuratons. The simplest explanation for this anomaly is probably that the method is not appropriate for such a small crystal section, where surface effects may dominate. Figure 3 shows surfaces of constant energy density for the $B = 4$ and the $B = 32$ configurations. In both cases, the energies have been chosen to highlight surface effects. It can be seen that rings of energy appear on the faces of the $B = 4$ cube where the crystal was cut. These are not seen in the true $B = 4$ solution; neither are they a feature of the crystal. They must therefore be an artifact of our ansatz. However, no such rings can be observed in the $B = 32$ configuration: in fact, there is very little distortion as the energy density surfaces go through the transition from the crystal to the outside region. It therefore appears that the model behaves quite well for $B = 32$ and larger $B$.

| $B$(calc) | $E^*$ | $L$ | $h$ | $n$ | $n/h$ |
|---|---|---|---|---|---|
| 4.000000 | 1.222 | 1.943 | 0.71 | 2.0 | 2.8 |
| 31.999997 | 1.193 | 2.108 | 1.84 | 4.8 | 2.6 |
| 107.999961 | 1.150 | 2.176 | 2.86 | 7.2 | 2.5 |
| 256.000108 | 1.125 | 2.215 | 3.87 | 9.6 | 2.5 |

Table 3: Detailed Results for Power Law Extrapolation: Simple Cubes.

| $B(calc)$ | $E^*$ | $L$ | $h$ | $\beta$ |
|---|---|---|---|---|
| 4.000000 | 1.189 | 1.857 | 0.61 | 1.89 |
| 32.000004 | 1.191 | 2.107 | 1.83 | 2.34 |
| 107.999954 | 1.149 | 2.176 | 2.86 | 2.34 |
| 256.000444 | 1.124 | 2.215 | 3.87 | 2.34 |

Table 4: Detailed Results for Exponential Extrapolation: Simple Cubes.

One other point about the cubic crystal sections is worth noting. The energy minima are extremely shallow with respect to both $h$ and either $\beta$ or $n$. To give an example, Figure 4 shows the energy of the $B = 108$ configuration for a variety of parameters. In fact, the energies had to be calculated more precisely than quoted in Tables 3 and 4, in order to pin down the values of the parameters.

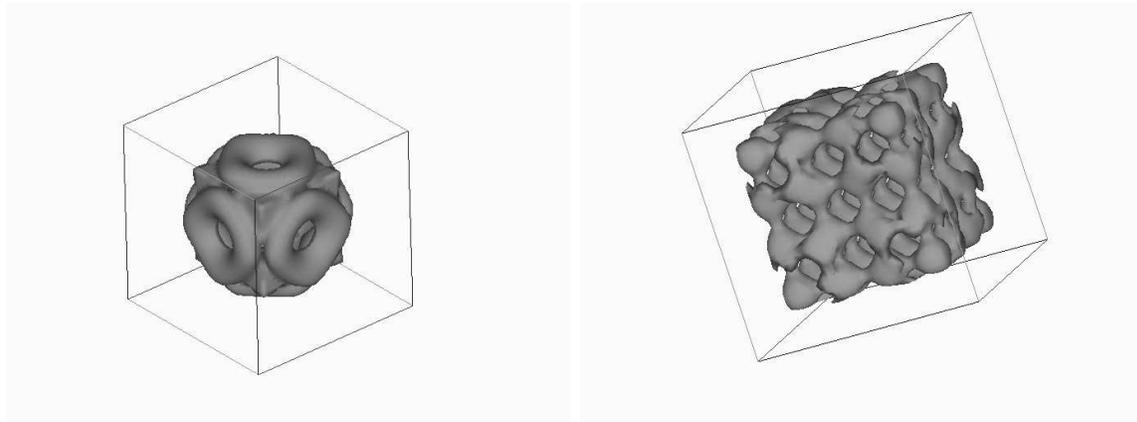

Figure 3: The B=4 (left) and B=32 (right) Configurations

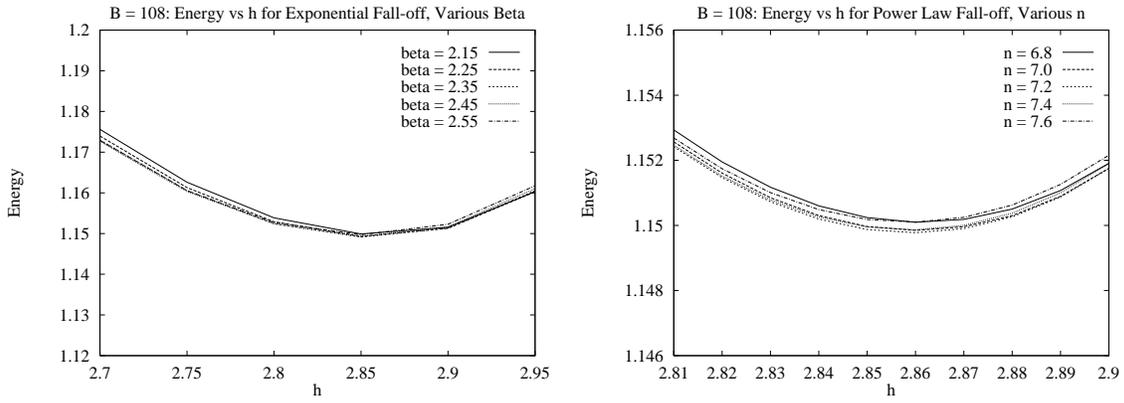

Figure 4: Variation of Parameters around Energy Minimum for B=108 (Both Extrapolations).

So far, we have managed to find three new, reasonably low energy Skyrme field configurations, which may be assumed to be approximations (giving an upper limit for the energy) of true minimal energy solutions. However, a major limitation of the model is that it is only possible to cut out sections corresponding to a very limited choice of baryon number. This may be partially remedied by expressing the energy formula (41) in terms of the baryon number:

$$E = 122.95\, B + 54.07\, B^{2/3} \quad \text{or} \quad E^* = 1.038 + 0.457\, B^{-1/3} \qquad (42)$$

Translating into physical units via Equation (5), and subtracting the skyrmion mass (864 MeV) from the volume term, we find that the binding energy as a function of $B$ is given by

$$E_B = 136B - 320B^{2/3} \text{ MeV}. \qquad (43)$$

This formula is probably best interpreted as the first two terms (or leading order effect) of a more general formula, approximating the semi-empirical mass formula of nuclear physics. However, this analogy should be treated with some caution. Classical Skyrme solutions cannot be directly compared to real nuclei. For a start, the skyrmion mass, subtracted to obtain a binding energy, differs significantly from the nucleon mass. It will be noted that the binding energies predicted by formula (43) are too high. This is typical of classical Skyrme solutions, where kinetic energy contributions are effectively ignored. These could well be of the order of 100 MeV [12]. The situation generally improves after quantisation. Also, our energy units, invoked in the conversion to MeV, depend on free parameters which are set to obtain agreement between experiment and *quantised* nucleon (and delta) solutions.

In fact, close examination of (43) reveals that for $B < 14$ the solutions are unbound. This may serve to give a quantitative estimate of the range in $B$ for which the model breaks down. It may also be observed that the surface term is rather high relative to the volume term, compared to the semi-empirical mass formula. This reflects the fact that most of the error in this model stems from the excess surface energy where the crystal is cut.

### 5.4 Other Shapes.

The value of cutting shapes other than simple cubes out of the crystal is rather dubious. Given that cubic symmetry must be maintained, simple cubes are the shapes which maximise the volume energy in comparison with the surface energy. Furthermore, many of the other possible shapes do not seem very likely for a nucleus; for example, some are elongated along the Cartesian axes. Even for those that are relatively feasible, it seems likely that the cubic formula (42) would give a better approximation to the energy of a real nucleus (or Skyrme solution) for the same baryon number.

One example will be given as a demonstration. The only two possibilities for $B < 100$ are $B = 28$ and $B = 76$, and we shall choose the former as our example. It consists of a central unit cell, with additional unit cells attached to each face. The generalisation of the outside ansatz required for this more complicated shape is very simple: radial extrapolations from *all* planes in one octant where $z$ is constant must be considered; as compared to a simple cube where there is only one such plane.

Although it is not possible to derive a general formula for the energies of the more complicated shapes, the quantities $V$ and $S$ given in Section 5.3 can still be used to make predictions for the volume plus surface energy in individual cases. For the $B = 28$ configuration, the prediction for $E^*$ is 1.244. This is to be compared with calculated values 1.261 and 1.262, obtained using an exponential and a power law fall-off respectively, outside the crystal boundary.

To reiterate the conclusions of the last section then, the calculation of the surface energy (per unit area) of the crystal, and the 'mass formula' derived from the result, are probably the most significant contributions of this method. The formula (42) gives an idea of what the Skyrme model predicts about volume and surface energies of the nucleus over a wide range of baryon numbers. The method also appears to give reasonably good approximations to cubically symmetric Skyrme solutions with $B = 32$, 108 and 256.

## Acknowledgments.

I would like to thank Dr. N. Manton for many useful discussions related to this work, and Dr. R. Leese for his help with the computing aspects of the project. In particular I would like to thank the latter for access to his codes and numerical routines, on which many of mine are based. I would also like to thank Dr. P. Sutcliffe for his help in producing the pictures which appear in Figure 3.